\documentclass[aps,pra,superscriptaddress,twocolumn,showpacs,noeprint]{revtex4-2}

\usepackage{graphicx,amsfonts,times,bm,amsmath,amssymb,verbatim,ifthen,braket,color,array,natbib,bm,soul}
\usepackage{array}
\usepackage{delarray}
\usepackage{xcolor}
\usepackage{subfigure}

\usepackage{xcolor} 

\newcommand{\be}{\begin{equation}}
\newcommand{\ee}{\end{equation}}
\newcommand{\bea}{\begin{eqnarray}}
\newcommand{\eea}{\end{eqnarray}}
\newcommand{\intz}{\mathbb{Z}}

\newcommand{\figfidelitymachinesz}{\ref{fig_fidelity_machines_z} }

\newcommand{\fighourglassgraph}{\ref{fig_hourglass_graph} }

\newcommand{\fighourglassexpgraph}{\ref{fig_hourglass_exp_graph} }

\usepackage{hyperref}

\begin{document}

\title{
Redundant String Symmetry-Based Error Correction: Demonstrations on Quantum Devices
}

\author{Zhangjie Qin}
\altaffiliation{These authors contributed equally to this work.}
\affiliation{Department of Physics, Virginia Tech, Blacksburg, Virginia 24061, USA}

\author{Daniel Azses}
\altaffiliation{These authors contributed equally to this work.}
\affiliation{School of Physics and Astronomy, Tel Aviv University, Tel Aviv 6997801, Israel}

\author{Eran Sela}
\affiliation{School of Physics and Astronomy, Tel Aviv University, Tel Aviv 6997801, Israel}

\author{Robert Raussendorf}
\affiliation{Institute for Theoretical Physics, Leibniz University Hannover, 30167 Hannover, Germany}

\author{V. W. Scarola}
\affiliation{Department of Physics, Virginia Tech, Blacksburg, Virginia 24061, USA}

\begin{abstract}
Computational power in measurement-based quantum computing stems from symmetry protected topological (SPT) order of entangled resource states.  But resource states are prone to preparation errors.  We introduce a quantum error correction approach using redundant non-local symmetry of the resource state. We demonstrate it within a teleportation protocol based on extending the $\mathbb{Z}_2 \times \mathbb{Z}_2$ symmetry of one-dimensional cluster states to other graph states.  Qubit \protect{ZZ-crosstalk} errors, which are prominent in quantum devices, degrade the teleportation fidelity of the usual cluster state. However, as we demonstrate on quantum hardware, once we grow graph states with redundant symmetry,  perfect teleportation fidelity is restored. 
We identify the underlying 
redundant-SPT order as error-protected degeneracies in the entanglement spectrum. 
\end{abstract}

\maketitle
\section{Introduction} Measurement-based quantum computation (MBQC) is carried out purely by  measurements of an entangled resource state \cite{raussendorf2001one,raussendorf2003measurement,briegel2001persistent}. Although MBQC 
does not use gates in the course of computation, gate errors affect resource state preparation.  Thus, implementing MBQC
in the current era of noisy quantum computers \cite{preskill2018quantumcomputingin,kitaev1997quantum,nielsen2010quantumcomputation,kitaev2003fault,calderbank1997quantum}, including numerous applications~\cite{PhysRevResearch.4.L032013,Zhang_2023,azses2023nonunitary},  requires quantum error correction (QEC) methods for MBQC.  A key source of errors in quantum devices that one would like to protect against is qubit ZZ-crosstalk 
~\cite{Dicarlo2009,McKay2019Three-QubitBenchmarking,Sarovar2020,Cai2021,Heinz2021,Kanaar2022,Xie_2022,Ni2022}.  Idle ZZ-crosstalk can arise, for example, from long range-interaction between qubits which are intrinsic to certain qubit platforms, e.g., Rydberg atom-based \cite{Saffman2010} resource state preparation.

The computational power of MBQC resource states relies on the presence of symmetries having a nontrivial action on the edge~\cite{else2012symmetry,miller2018latent,miller2016hierarchy,miyake2010quantum,raussendorf2019computationally,stephen2019subsystem,Azses_2023}, i.e. on symmetry protected topological (SPT) order \cite{wen2017colloquium,chen2010local,chen2013symmetry,chen2011complete,kitaev2001unpaired,moore2007topological,oshikawa1992hidden,perez2008string,pollmann2012symmetry,ryu2010topological,senthil2015symmetry}. SPT phases can be detected by degeneracies in the entanglement spectrum or their related string order parameters (SOPs)~\cite{pollmann2010entanglement,turner2011topological,choo2018measurement,cornfeld2019entanglement,azses2020symmetry,Azses_2023}. In this work, we apply such a symmetry-based approach to propose a QEC method in MBQC protocols. Generally, SPT states are characterized by a certain symmetry group. As the main example on which we focus, Ref.~\cite{else2012symmetry,son2011topological} proved that 1D SPT states with $(\intz_2)^2$ symmetry \cite{son2011topological} are resource states for 1D MBQC.  Therefore, MBQC is immune to symmetry-preserving noises \cite{azses2020identification}. However, many common noise sources are not symmetric, e.g., ZZ-crosstalk as discussed below. 

Here we introduce an approach to make MBQC protocols error-oblivious by growing graph states with extended symmetry groups, $(\intz_2)^g$, where $g$ is a graph-dependent integer.  Our approach gives a framework to understand existing error oblivious graph states~\cite{Morley-Short2019} and to construct new ones. We focus on teleportation fidelity as a non-local probe, related to tests of computational power based on the SOP~\cite{RAUSSENDORF2023,raussendorf2017symmetry,miller2015resource}. 

\begin{figure}[t]
\centering
(a){\includegraphics[width=0.4\linewidth]{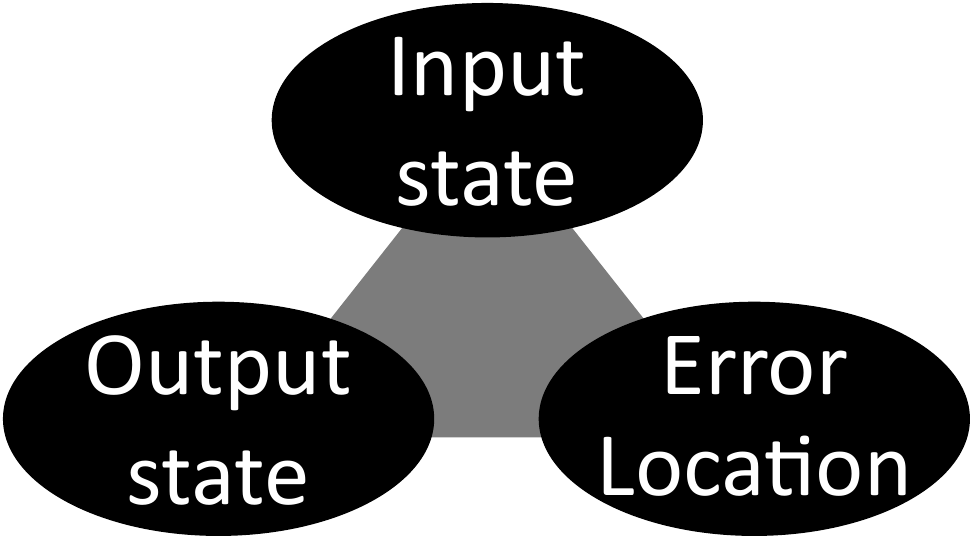}}
(b){\includegraphics[width=0.5\linewidth]{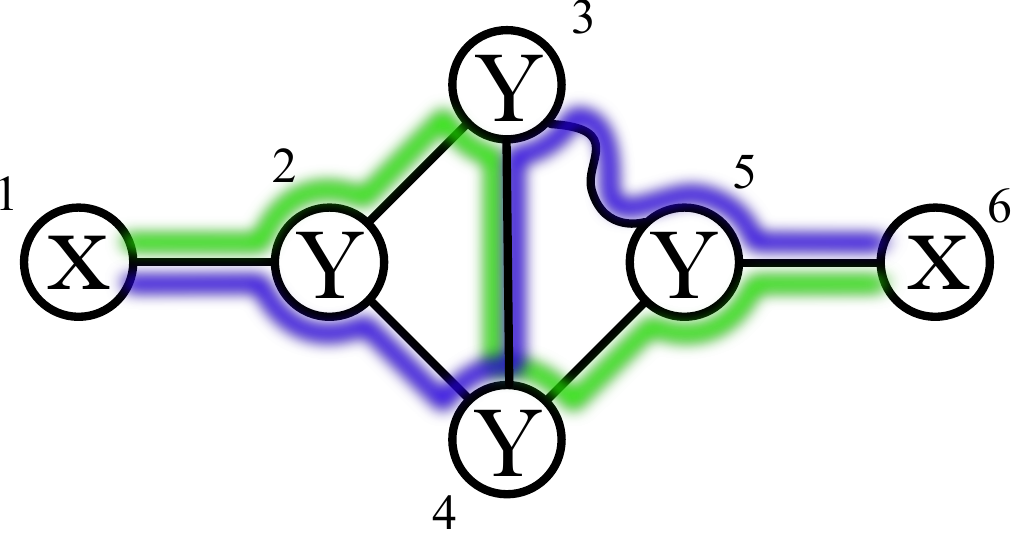}}
\caption{\label{fig_diamond_path_schematic}
(a)  \emph{Decision Triangle}: If any two items are known, our protocol reveals the third.  (b) \emph{Diamond Graph}: The simplest example of a graph state with redundant symmetry. The vertices are marked with $X$ or $Y$ depending on their stabilizers, see Eqs.~(\ref{stab:x}) and (\ref{stab:y}), respectively. The different paths, marked with blue and green, show the $(\intz_2)^2 \times (\intz_2)^2$ symmetry, where each path contributes $(\intz_2)^2$ to this redundant symmetry form. The wavy line between vertices 3 and 5 represents an idle ZZ-crosstalk error. The input (output) $I$ ($O$) is the left (right) most vertex.
}
\label{fig_logic_diamond_path}
\end{figure}

MBQC based teleportation corresponds to a path from input to output, along which errors may occur. As displayed in Fig.~\ref{fig_diamond_path_schematic}a, our error correction methods allow us to deal with errors.  Knowledge of the input and output states identifies the error. Likewise, knowledge of one of the input or output states and the error, allows us to deduce the other state. 

Our symmetry redundancy based error protection can be understood in terms of multiple teleportation paths. The various errors that may occur shrink the symmetry group, but as long as a minimal amount of symmetry persists, teleportation is unaffected. This corresponds to a particular path connecting input and output. After introducing the stabilizer formalism, we make this picture explicit in various graph states, see for example Fig.~\ref{fig_diamond_path_schematic}b. We demonstrate our construction using different numerical methods and real noisy quantum computers (IonQ and IBMQ) where we consider protection against ZZ-crosstalk and single qubit errors.

The paper is organized as follows. In Sec.~\ref{se:1} we introduce graph states, which form the basis for MBQC, as well as graph states perturbed by errors. In Sec.~\ref{se:2} we provide a simple exposition to MBQC and teleportation in a  1D graph, and discuss its limited robustness against errors. Then, in Sec.~\ref{se:3} we formalize our  general idea of enhancing the symmetry of the graph state as a method for error protection. 
We provide results on real quantum computers in Sec.~\ref{se:results} and in Sec.~\ref{se:4}  we further exemplify this concept in the context of a particular graph state~\cite{Morley-Short2019} with higher symmetry. In Sec.~\ref{se:telo_pert} we apply the formalism to ground states of perturbed stabilizer Hamiltonians. Finally, in Sec.~\ref{app:reduced_dm}
we show that the error protection in these various graphs results from error protected degeneracies in the entanglement spectrum. We summarize in Sec.~\ref{se:5}.

\section{Error-prone graph states} 
\label{se:1}
Graph states are stabilizer states corresponding to a graph.
Consider a graph $G=(V,L)$ consisting  of vertices $i \in V$ and links $(i,j) \in L$.
Conventional graph states $\ket{\psi_G}$ are the unique eigenstates of the stabilizer elements \cite{gottesman1997stabilizer} 
\be
\label{stab:x}
K_i= X_i\prod_{j\in \mathcal{N}(i)} Z_j,
\ee
where $\mathcal{N}(i)$ is the neighborhood of vertex $i$ connected to it by a link. $X_i$ and $Z_i$ are Pauli matrices acting at vertex $i$.  To create the state stabilized by the $K_i$'s one starts from the product state $\prod_{i} \ket{+}_i$ where $|+\rangle=(|0\rangle + |1 \rangle)/\sqrt{2}$, and then entangles it by applying  $CZ_{ij} = e^{-i \frac{\pi}{4} (I-Z_i-Z_j+Z_i Z_j)}$ gates on all links.

Here we are concerned with the preparation of graph states on quantum computers \cite{hein2006entanglement,schwartz2016deterministic}.  We assume that, in addition to  single qubit rotations, we have access only to native Ising gates \be
\label{eq_Ising_gate}
U_{ij}^{ZZ} (\epsilon_{ij})= e^{-i (\pi/4+\epsilon_{i,j}) Z_i Z_j}=e^{-i (\pi/4) Z_i Z_j} E(\epsilon_{ij}), 
\ee
where $E(\epsilon_{ij})=e^{-i \epsilon_{i,j} Z_i Z_j}$ is an idle ZZ-crosstalk error \cite{Sarovar2020} parameterized by $\epsilon_{i,j}$ that may vary from link-to-link.   

Thus, we consider states of the form 
\be
\vert \Psi_G (\{\epsilon_{i,j} \})\rangle\equiv \prod_{i,j \in L} U_{ij}^{ZZ}(\epsilon_{i,j}) \left[ \prod_{i\in V} e^{i\theta_i Z_i} \vert +\rangle_i \right].\label{eq_Ising_graph_state}
\ee
The $\theta_i$'s are specified for each graph as exemplified below, such that ideal graph states, $\vert \Psi_G (\{\epsilon_{i,j} =0\})\rangle$, are stabilizer states with a minor modification: some vertices contain an additional $e^{-i \frac{\pi}{4}Z_i}$ rotation, thus the stabilizers at these vertices are replaced by 
\be
\label{stab:y}
K_i= Y_i\prod_{j\in \mathcal{N}(i)} Z_j.
\ee
We mark those vertices in the figures by $Y$ instead of $X$. We note that this choice does not affect the commutativity of stabilizers. The motivation for this choice will become clear below, as it simplifies the teleportation circuit. 

\begin{figure}
\begin{center}
\includegraphics[width=0.4\textwidth]{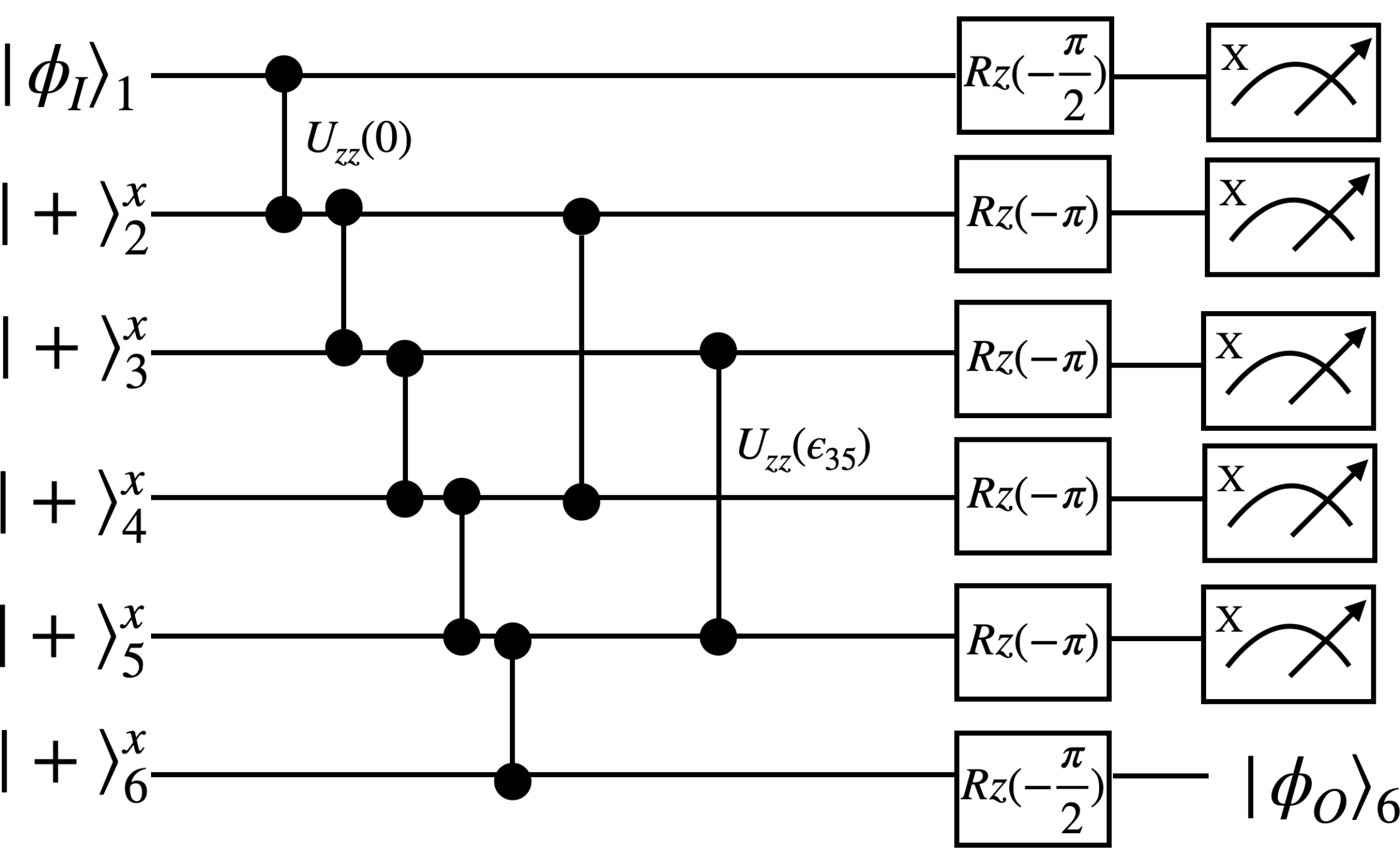}
\end{center}
\caption{\emph{Diamond Graph Circuit Diagram}:  Gate sequences used to obtain data shown in Fig.~\figfidelitymachinesz (top).   All but qubit 1 are first aligned along the $x$-direction.  Qubit 1 is prepared in the state $\phi_I$.  The qubits are then entangled with Ising gates.  The gate entangling qubits 3 and 5 are perturbed by $\epsilon_{35}$.  Single qubit $z$-rotations are then applied, where $R_{z}(\gamma) =e^{-i\gamma Z/2}$.  Measurements of all but qubit 6 along the qubit-$x$ direction propagates the state from qubit 1 to qubit 6.  Qubit 6 is measured in the basis of $\phi_I$ to construct the fidelity. }
\label{fig_circuit_diamond}
\end{figure}

Fig.~\ref{fig_diamond_path_schematic}b shows an example, the diamond graph.  It contains an input and output vertex, 1 and 6, respectively.  It is the simplest example with redundant paths. Its circuit realization is shown in Fig.~\ref{fig_circuit_diamond} where the angles $\theta_i$ are denoted explicitly.

\section{SPT phase and teleportation} 
\label{se:2}
To test if the teleportation property of a resource state $\ket{\psi_G}$ is protected against a unitary perturbation such as $E(\epsilon_{ij})$  we first introduce the symmetries $\{ s_\alpha\}$ which generate the symmetry group of the graph, $\mathcal{S}$, i.e. $s_\alpha \ket{\psi_G} = 
\ket{\psi_G}$ for any $s_\alpha \in \mathcal{S}$.  Each symmetry  is a Pauli string generated by the stabilizers \begin{align}
s_\alpha\equiv \prod_{j\in \alpha} K_j,
\label{eq_general_chain}
\end{align}
where $\alpha$ denotes a subset of the vertices which can be non-local.  Each symmetry can be factorized into three parts, acting on the input, middle, and output, as $s_\alpha=s_\alpha^{(I)} s_\alpha^{(M)} s_\alpha^{(O)}$. By measuring the middle region in the Pauli basis according to $s_\alpha^{(M)}$, the operator is transformed to a  $\pm 1$ sign depending on the measurement results. We assume that a simultaneous measurement exists, implying that $s_\alpha^{(M)}$ commute for different $\alpha$.  Finally there should be at least two such symmetries, such that the common eigenstate of both  $s_\alpha^{(I)} s_\alpha^{(O)}$ $(\alpha = x,z)$ is a maximally entangled Bell state between the input and output. 

This can be exemplified for the 1D chain graph. Consider the state  stabilized by the $X$ or $Y$ stabilizers in Eqs.~(\ref{stab:x}) and (\ref{stab:y}), as denoted in  Fig.~\ref{fig_diamond_path_schematic}b, but with the 2-4 and 3-5 bonds disconnected, i.e. only with nearest neighbors.  It has an odd and even symmetry generators $s_{\mathrm{odd}} = s_{135} = X_1 Y_3 Y_5 Z_6$ and $s_{\mathrm{even}} = s_{246} = Z_1 Y_2 Y_4 X_6$, thus $\mathcal{S}=\mathbb{Z}_2 \times \mathbb{Z}_2$. All the $s_\alpha^{(M)}$ involve only Pauli-$Y$ operators hence they commute. By measuring all middle qubits in the $x$-basis we obtain a Bell state between the input and output qubits. Therefore, this state is a resource state for general 1-qubit MBQC. As in this example, one can always write~\cite{raussendorf2003measurement} $s_x^{(I)}=X_I$, $s_z^{(I)}=Z_I$ and $s_x^{(O)}=U(X_O)$, $s_z^{(O)}=U(Z_O)$ for some unitary $U$ which does not change the entanglement of the Bell state, where $U(O)=UOU^\dagger$.

Consider an error described by a unitary operator $E$ affecting the resource state, $\ket{\Psi_G'} = E \ket{\Psi_G}$. As long as $E$ commutes with $\mathcal{S}$, it does not alter the relation $s_\alpha|\Psi_G'\rangle=|\Psi_G'\rangle$, and hence the perfect teleportation property persists \cite{azses2020identification}.  However, perturbations such as $E_{ij}^{ZZ}(\epsilon_{i,i+1})$ on the chain graph state do not commute with these symmetries in general (unless the idle ZZ-crosstalk acts only on even spaced vertices $j-i = 2m$, where $m \in \mathbb{N}$).  Is it possible to restore perfect teleportation in the presence of such non-symmetric perturbations? 

\section{Redundant symmetry} 
\label{se:3}
Our approach is to work with a graph having an extended symmetry group. Starting from such a higher symmetry, consider the symmetry subgroup $\mathcal{S}^{(E)}$ which commutes with a perturbation $E$, i.e. $[E,s_\alpha]=0$ for any $s_\alpha \in \mathcal{S}^{(E)}$. If this subgroup contains $(\intz_2)^2$, then computational power persists in the presence of this perturbation. This protection holds for different perturbations $E_i$ that may happen one at a time, even if $\mathcal{S}^{(E_i)}$ are different, as long as they all contain $(\intz_2)^2$.  Furthermore, these sub-group symmetries generate a nontrivial redundancy if they share one or more vertices besides $I$ and $O$.  Many sub-group symmetries can be found to commute for Ising-based graphs [Eq.~\eqref{eq_Ising_graph_state}] since link operators of the form $\{I_{i} I_{j}, X_{i}X_{j},Y_{i}Y_{j},Z_{i} Z_{j}\}$ all commute on the same link.

\begin{figure}
\centering
\includegraphics[width=\linewidth]{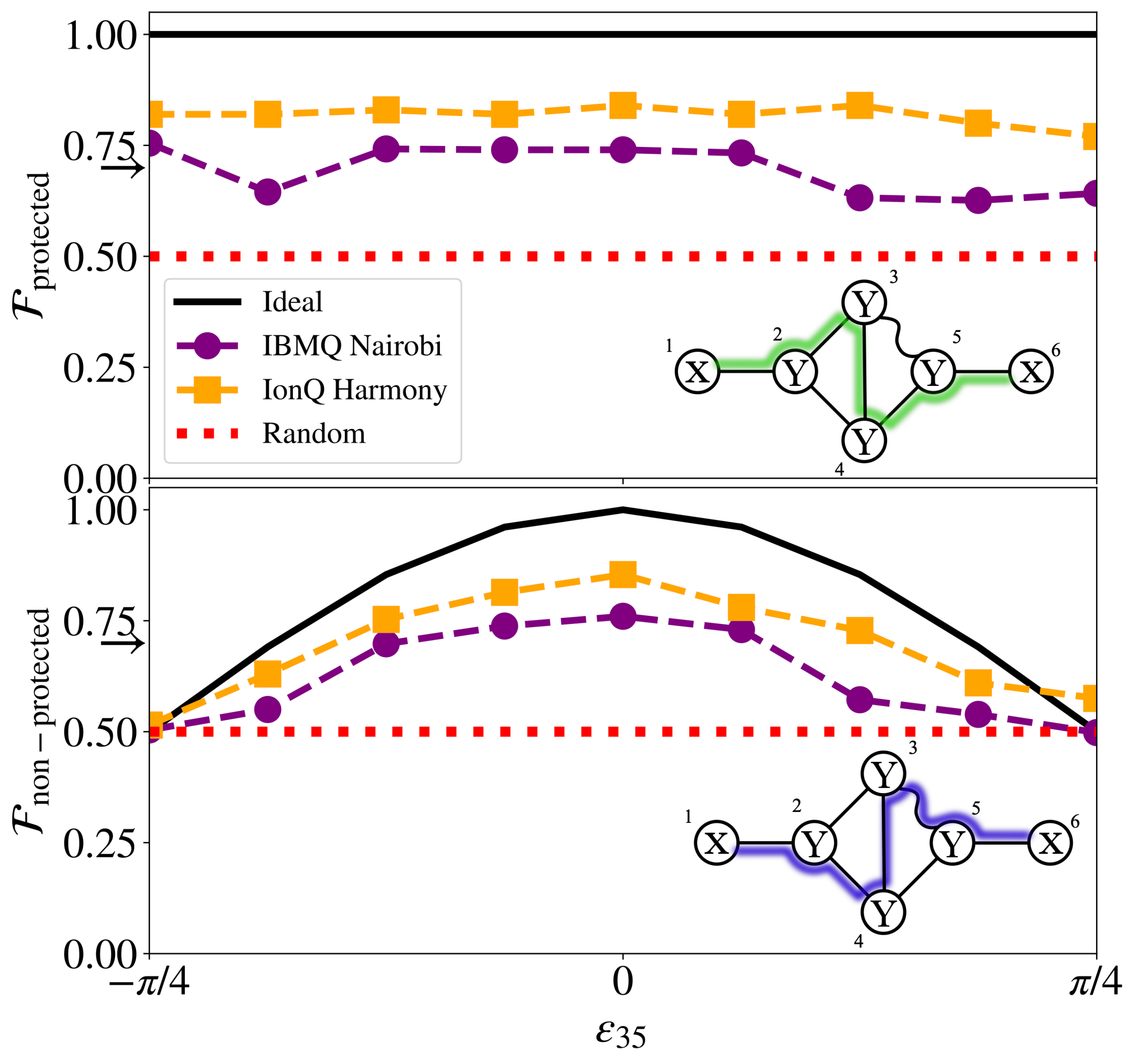}
\caption{
Fidelity for teleportation in the diamond graph along a preserved (top) and broken (bottom) symmetry path plotted as a function of the perturbation strength to the native Ising gates.  The solid line in the top panel indicates perfect transmission.  Square (Circle) data are obtained from IonQ (IBMQ) quantum devices.  These data are lower than unity due to noise sources unprotected by topological symmetry, e.g., two-qubit depolarizing noise, see Appendix~\ref{app:circuit_device}. The arrow at 2/3 shows the threshold above which transmission along a quantum channel is guaranteed.  The dotted line at 1/2 indicates a random classical channel.
}
\label{fig_fidelity_machines_z}
\end{figure}

To demonstrate symmetry-redundant QEC consider the diamond graph with a preparation ZZ-crosstalk error, e.g., between qubits 3 and 5 as marked by a wavy line in Fig.~\ref{fig_diamond_path_schematic}b. The unperturbed graph state has a higher symmetry compared to the 1D chain graph. This can be understood geometrically by considering two entwined paths connecting the input and output qubits, (green and blue paths in Fig.~\ref{fig_diamond_path_schematic}b). 
The blue path allows introduction of one pair of even and odd symmetries,
$s_{135}=X_1 X_3 X_5 Z_6$ and $s_{246}=Z_1 X_2 X_4 X_6$. However, the green path allows introduction of another pair of even and odd symmetries, $s_{145}=X_1 X_4 X_5 Z_6$ and $s_{236}=Z_1 X_2 X_3 X_6$. We can see that all the middle sequences involve only $X_i$'s, This was achieved due to our incorporation of the unconventional $Y-$stabilizers in Eq.~(\ref{stab:y}), and will allow to teleport using $X$-measurements only as in Fig.~\ref{fig_circuit_diamond}. 

The total symmetry group is thus $(\mathbb{Z}_2)^4$. Let us  label the symmetries as $(p,q)$ according to the two symmetries starting with $q=x,z$ corresponding to each path $p$,
\bea
\label{eq:spi}
s_{p,x}&=&X_I s^{(M)}_{p,x} U(X_O), \nonumber \\
s_{p,z}&=&Z_I s^{(M)}_{p,z} U(Z_O), 
\eea
where we have $I=1$, $O=6$, $U$ is the Hadamard matrix, $s_{1,x}=s_{135}$, $s_{1,z}=s_{246}$; $s_{2,x}=s_{145}$ and $s_{2,z}=s_{236}$.

The perturbation $\epsilon_{i,j}$ commutes only with the $p=2$ [
$p=1$] symmetries  for crosstalk on links $(i,i+2)$ for $i=2,3$ [$(i,i+1)$ for $i=2,4$].  For example, the $\epsilon_{35}$ error  in Fig.~\ref{fig_diamond_path_schematic}b commutes only with 
$\{s_{135},s_{246} \}$, i.e. the symmetries corresponding to the green path $p=1$.  This is sufficient to guarantee  perfect teleportation fidelity. In the next section, we  demonstrate this oblivious teleportation using $\mathbb{Z}_2^4$ SPT order on a  quantum machine.

\begin{figure}[t]
\centering
\includegraphics[width=0.7\linewidth]{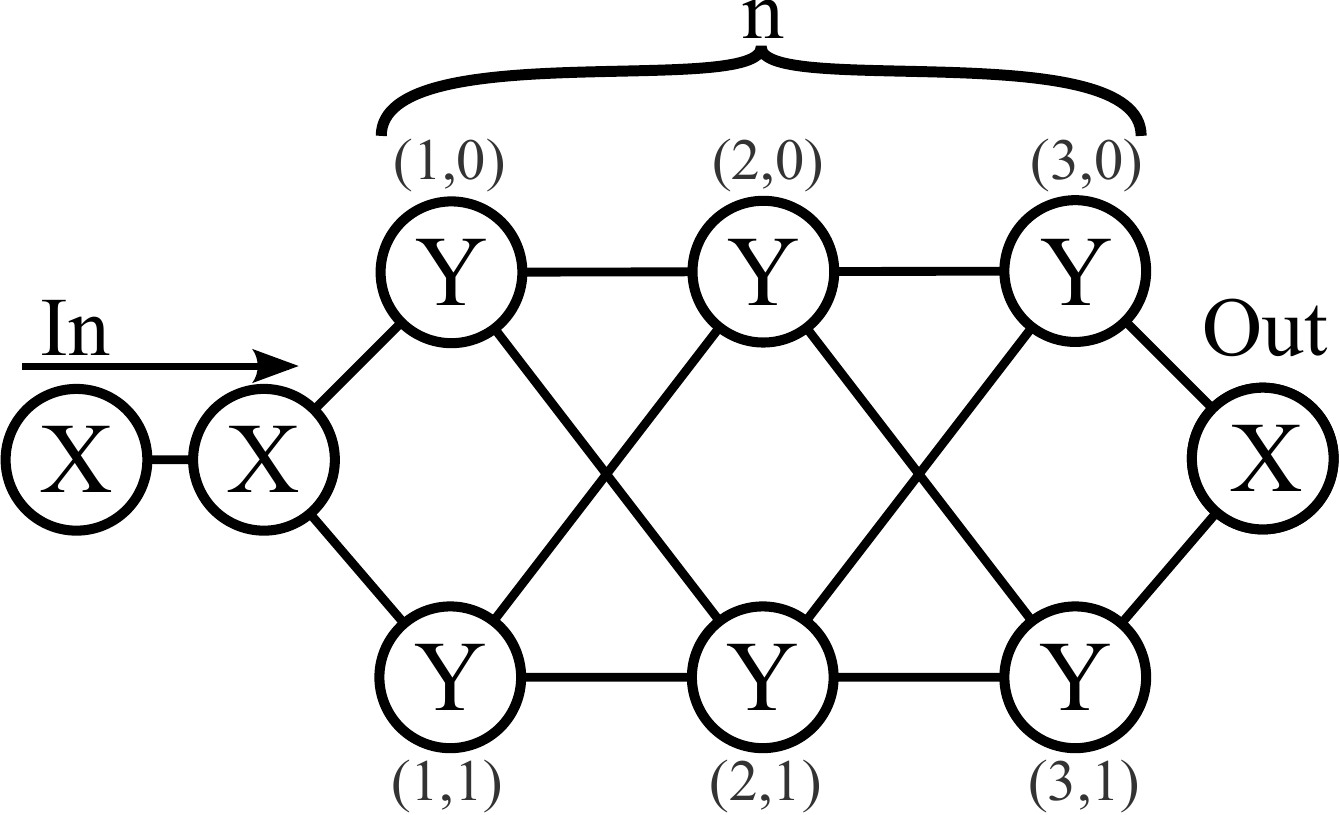}
\caption{\label{fig_hourglass_graph}\emph{Hourglass Graph}~\cite{Morley-Short2019} containing $N=2n+3$ qubits. It contains $2^n$ paths connecting input and output.
}
\end{figure}

\section{Results}
\label{se:results}
To demonstrate teleportation robustness, we prepare qubit 1 in state $\vert \phi_I^{\hat{r}}\rangle$, where the unit vector $\hat{r}$ corresponds to a location on the Bloch sphere.  Quantum teleportation can be seen as an MBQC identity gate since the output state is the same as the input $\phi_O^{\hat{r}} = \phi_I^{\hat{r}}$.  Single-qubit measurements on input and middle qubits (1-5 in Fig.~\ref{fig_diamond_path_schematic}b) teleport information from the graph input to the output \cite{raussendorf2001one,raussendorf2003measurement,Tame2005,TAME2006,Qin2021}. Let us denote by  $s_i=0,1$ the measurement outcomes of each of these 5 measurements.  The success of the teleportation is measured by the fidelity $\mathcal{F}^{(p)}_{\hat{r}}\equiv\vert \langle \phi_I^{\hat{r}} \vert U^{(p)}_{\Sigma}(\{ s_i\}) \vert \phi_O^{\hat{r}} \rangle \vert^2$, which compares the input qubit state $\phi_I^{\hat{r}}$ to the output qubit state $\phi_O^{\hat{r}}$.  Here, $U_{\Sigma}^{(p)}$ is a path dependent byproduct operator. In general MBQC, the byproduct operators describe the dependence of the resulting gate, on measurement outcomes~\cite{raussendorf2003measurement}. 

In our case,  let us consider separately teleportation across the green and blue paths in \ref{fig_diamond_path_schematic}b (denoted $p=1,2$, respectively). For the green path $(p=1)$ we find
\begin{align}
    U_{\Sigma}^{(p=1)}(\{ s_i\}) &=(Z)^{s_2+s_4}(X)^{s_1+s_3+s_5}.   
\end{align}
Namely, as long as $s_{135}$ and $s_{246}$ are symmetries, the output state obtained after 5 measurements with results $s_i$ is $|\phi_O \rangle=OU_{\Sigma}^{(p=1)}(\{ s_i\}) |\phi_I \rangle$ for any incoming state.  $O$ is the Hadamard (Identity) matrix for paths with an even (odd) number of vertices on the path. This can be shown using usual MBQC methods~\cite{raussendorf2003measurement}. Similarly, using the blue path, we obtain another byproduct operator,
\begin{align}
   U_{\Sigma}^{(p=2)}(\{ s_i\})&=(Z)^{s_2+s_3}(X)^{s_1+s_4+s_5}.  
\end{align}
Next we use these two byproduct operators to calculate the fidelity.  We present quantum device results for $\mathcal{F}=\mathcal{F}_{\hat{z}}$. As discussed in 
Appendix~\ref{app:mps_non_symmetric1}
we obtained the same qualitative results for other $\hat{r}$. 

Figure~\ref{fig_fidelity_machines_z} plots the fidelity for two quantum devices. In the top panel we apply the byproduct operators of the green path, and in the bottom panel we apply the byproduct operators of the blue path. Ideally we have perfect teleportation, $\mathcal{F}=1$, as shown in the top panel by a black line.  We can see that for both quantum devices the teleportation is not perfect due to noise which is excluded in our symmetry analysis. Nonetheless, the graph state entanglement is ensured to allow transmission across a quantum channel \cite{Horodecki1999,Braunstein2001,Bose2003,Paternostro2005,Christandl2005} if $\mathcal{F}_{\hat{r}}>2/3$. 
We can clearly see that the ability to teleport survives in the presence of the $\epsilon_{35}$ only if we teleport along the green path. The effects of uncontrolled noises of the quantum devices can be accounted for.  In Appendix~\ref{app:circuit_device} we reproduce the observed fidelity trend for the IonQ device using a two-qubit depolarizing noise model.  

\begin{figure}[t]
\centering
\includegraphics[width=\linewidth]{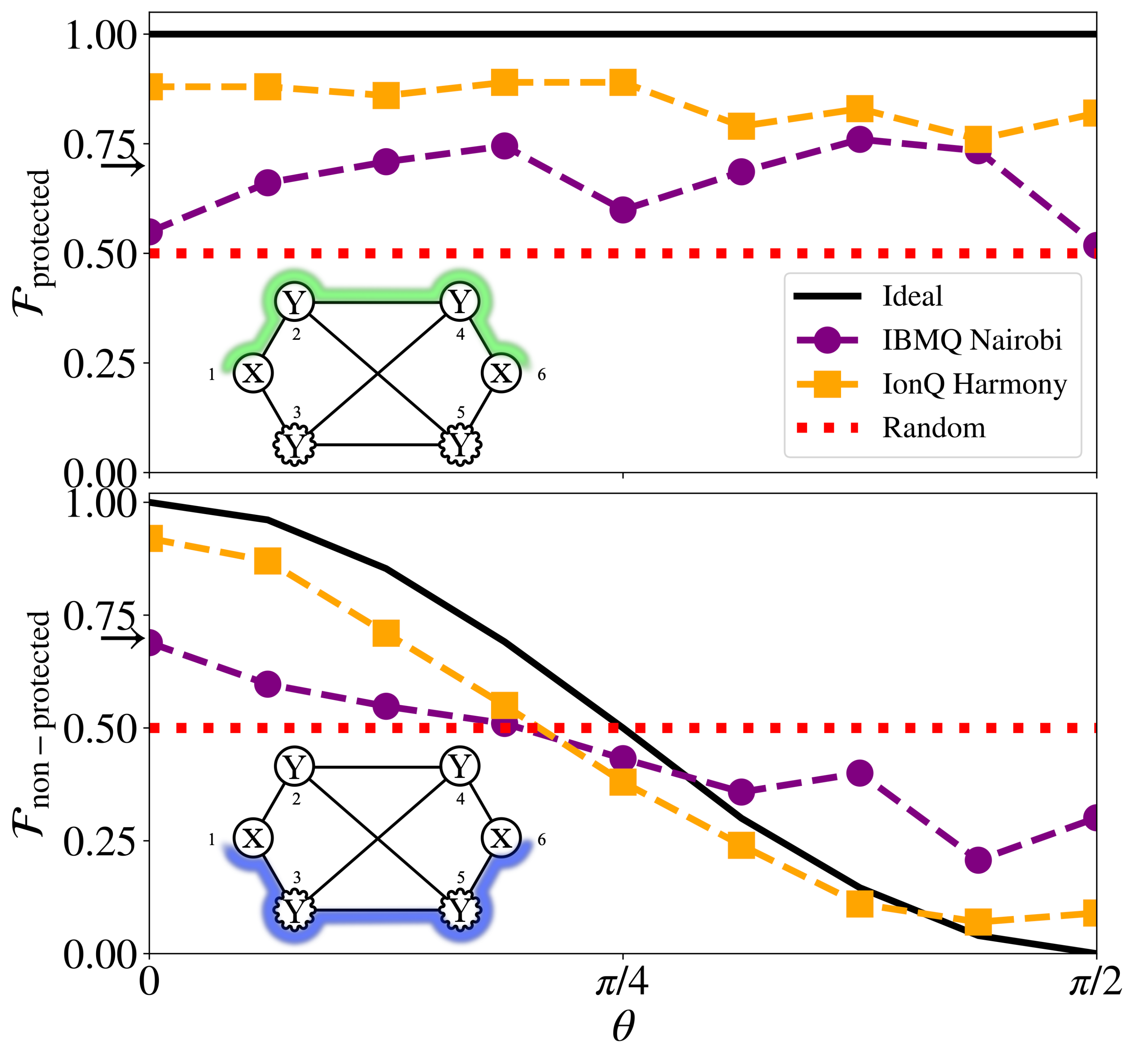}
\caption{
The same as Fig.~\ref{fig_fidelity_machines_z} but for the hourglass graph with $n=2$ and single qubit errors, $e^{-i\theta (X_{3}+X_{5})/2}$, indicated by wavy lines on bottom vertices.  The top panel shows evidence of teleportation across a quantum channel protected by the upper non-local string symmetry in both quantum devices.
}
\label{fig_hourglass_exp_graph} 
\end{figure}

Here we assumed that only one error (crosstalk event) took place. In the case of simultaneous crosstalk on both paths, the redundant-SPT order does not protect the information flow. As discussed above, in MBQC one has to apply byproduct operators depending on the measurement outcomes, and on the path. Thus, one needs to know the path through which the information was transmitted. This can be done by a calibration process  according to error-tomography with the decision triangle, Fig.~\ref{fig_logic_diamond_path}a.  

The diamond graph is a simple illustration of teleportation of quantum information over redundantly many paths in the presence of errors. It is highly limited to specific errors. For example, it can not correct errors on any link. Indeed, both green and blue paths are corrupted for an error of the form $\epsilon_{34}$. Similarly, the diamond graph is limited to ZZ, XX or YY link errors, and can not deal, e.g., with XY errors. Below we discuss a generalization that deals with more general errors.

\section{Hourglass Graph}
\label{se:4}
One possible generalization of the diamond graph to many qubits is the hourglass graph shown in Fig.~\ref{fig_hourglass_graph} and studied in Ref.~\cite{Morley-Short2019}. Its quantum circuit is shown in Appendix~\ref{app:circuit_device}. 
As we will show, it protects MBQC teleportation from any unitary perturbation in several bulk links or vertices.  As in the diamond graph, one has several routes of information from input to output. For example the hourglass graph with odd $n$ has the following upper path $(k=0)$ and lower path $(k=1)$ symmetries,
\bea
\label{hourglasssym}
X_m Y_{2,k} Y_{4,k} \dots X_O, \nonumber \\
Z_m Y_{1,k} Y_{3,k} \dots Z_O .
\eea
However, in this case the number of paths grows exponentially with the linear size $n$. Indeed the operators $X_m Y_{2,k_2} Y_{4,k_4} ,\dots,Z_O$ and $Z_m Y_{1,k_1} Y_{3,k_3} ,\dots,Z_O$ are symmetries for any $k_j=0,1$. For each additional two vertices in the bulk, the number of paths multiplies to yield $2^{n}$ possible paths.  In this graph, 1-qubit errors in the middle region reduces the number of uncorrupted paths to half, i.e. we still can teleport quantum information via $2^{n-1}$, and at least one uncorrupted path can be found for anytype of two-qubit error. 

As for the diamond graph, we have to apply different byproduct operators in order to teleport quantum information along different paths. For an even linear size $n$, if we measure the input qubit in the $x$-direction and bulk qubits all in the $-y$ direction, the condition $|\phi_O^{\vec{r}} \rangle= OU_{\Sigma}^\text{HG}|\phi_I^{\vec{r}} \rangle$ defines the byproduct operator.  There are $2^n$ paths we can select. When $n$ is an even number, we choose the pair of paths shown in the inset of Fig.~\fighourglassexpgraph and label them as upper and lower. The corresponding upper and lower path byproduct operators for $n$ even are given by
\begin{align}
    U_{\Sigma}^{\text{(u)}}&=(Z)^{s_{in}+s_{(1,0)}+\dots+s_{(n-1,0)}}   (X)^{s_m+s_{(2,0)}+\dots+s_{(n,0)}},
\end{align}
and
\begin{align}
    U_{\Sigma}^{\text{(l)}} &=(Z)^{s_{in}+s_{(1,1)}+\dots+s_{(n-1,1)}}    (X)^{s_m+s_{(2,1)}+\dots+s_{(n,1)}},
\end{align}
respectively.  Here $s_{(i,j)}$ labels measurement outcomes on bulk vertex $(i,j)$ in Fig.~\fighourglassgraph and $s_m$ labels the vertex to the right of the input qubit $S_{in}$ and to the left of $s_{1,i}$. Fig.~\ref{fig_hourglass_exp_graph} shows the fidelity for the hourglass graph state with $n=2$, for errors applied to the marked lower vertices. We compute the fidelity either based on the upper path (upper panel) or lower path (lower panel). A comparison of top and bottom panels shows that the upper path allows teleportation along a quantum channel for both quantum devices (top panel with $\mathcal{F}>2/3$ ). 

In practice, one needs to find the broken link in order to select the correct byproduct operators. To know which path has not been corrupted by an error, we construct a post-processing calibration protocol in Appendix~\ref{app:mbqc_path_calibration}.

\section{Teleportation through perturbed ground states}
\label{se:telo_pert}

\begin{figure}
\begin{center}
\includegraphics[width=0.8\linewidth]{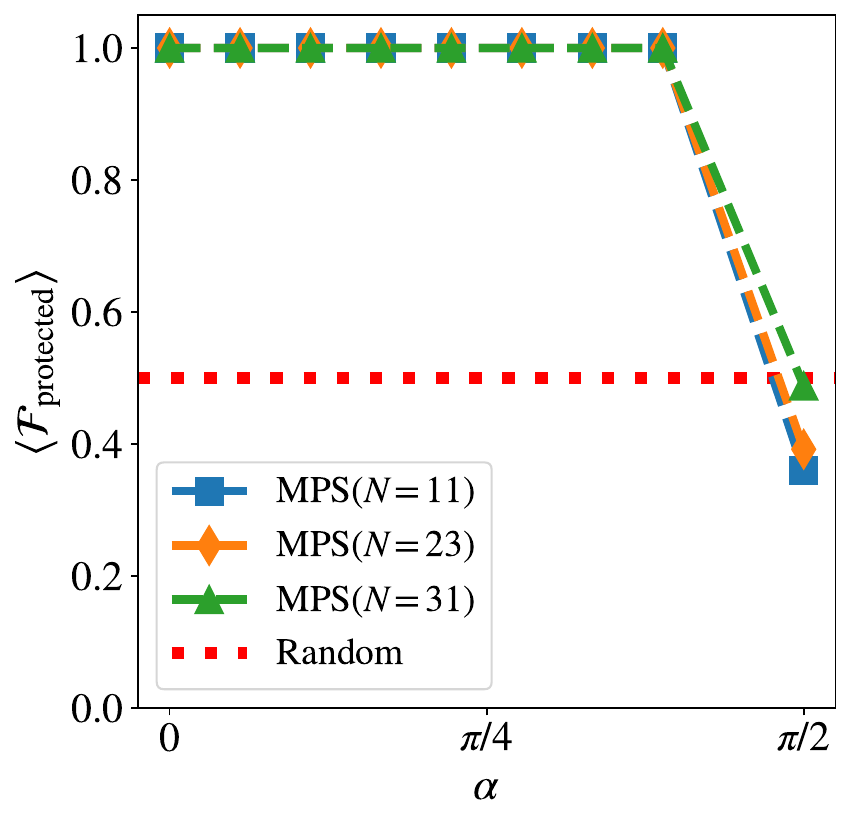}
\end{center}
\caption{\label{fig:fig_fidelity_up_sym} MBQC teleportation fidelity for the upper path in the hourglass graph perturbed as in $H_{\text{Y}}$. The calculations were done using MPS 
for 25 random input states, for further details see Appendix~\ref{app:mps_symmetric}. The results show the protection of the teleportation protocol for $\alpha \neq \pi/2$ along the upper path.  Due to convergence problems of the bond dimension, we excluded one point near $\pi/2$.
}
\end{figure}

So far, we applied the teleportation protocol on perturbed stabilizer states  within the circuit model, for example by applying ZZ crosstalks in Fig.~\ref{fig_circuit_diamond}. In this section, we illustrate our formalism of multi-path teleportation, on ground states of perturbed Hamiltonians, which may not be  easy to prepare on a quantum computer. We focus on three cases, in each of which we add  a specific perturbation on top of a Hamiltonian $H=-\sum_{j=1}^L K_j$, where $K_j$ correspond to either $X$ or $Y$ stabilizers as denoted in  the hourglass graph in Fig.~\ref{fig_hourglass_graph}. We   first ignore the  input qubit, which will be incorporated below. We notice that the unperturbed stabilizer state has the symmetries of the form of Eq.~(\ref{hourglasssym}). Below, we add three types of perturbations: (i) symmetric perturbation, i.e. and operator which commutes with all the symmetries, (ii) non-symmetric perturbation acting only on the lower path, and  (iii) non-symmetric perturbation acting on one path only. In each case we study the robustness of teleportation fidelity.  To deal with a large linear size $n$ we apply exact diagonalization and matrix product state (MPS) simulations \cite{White1992,schuch2011classifying,Dolfi2012,Orus2014,Bridgeman2017,biamonte2017tensor,Fishman2022}.  

\subsection{Symmetric perturbation}
We perturb the stabilizer Hamiltonian by adding the following symmetry preserving terms
\bea
\label{eq:Hy}
H_{\text{Y}}= -\cos{\alpha}   \sum_{j=1}^{L} K_j  -\sin{\alpha}  \sum_{i=1}^n (Y_{i,0}+Y_{i,1})  ,
\eea 
where $L=2n+2$ is the number of stabilizer terms.  As detailed in Appendix~\ref{app:mps_symmetric}, after finding the ground state of this Hamiltonian using an MPS approximation, the input qubit is entangled with the first site on the left, denoted $m$, and subsequently measurements are performed in order to realize our teleportation protocol along a desired path. We also define the total number of qubits to be $N=L+1$.

 \begin{figure}
\begin{center}
\includegraphics[width=.8\linewidth]{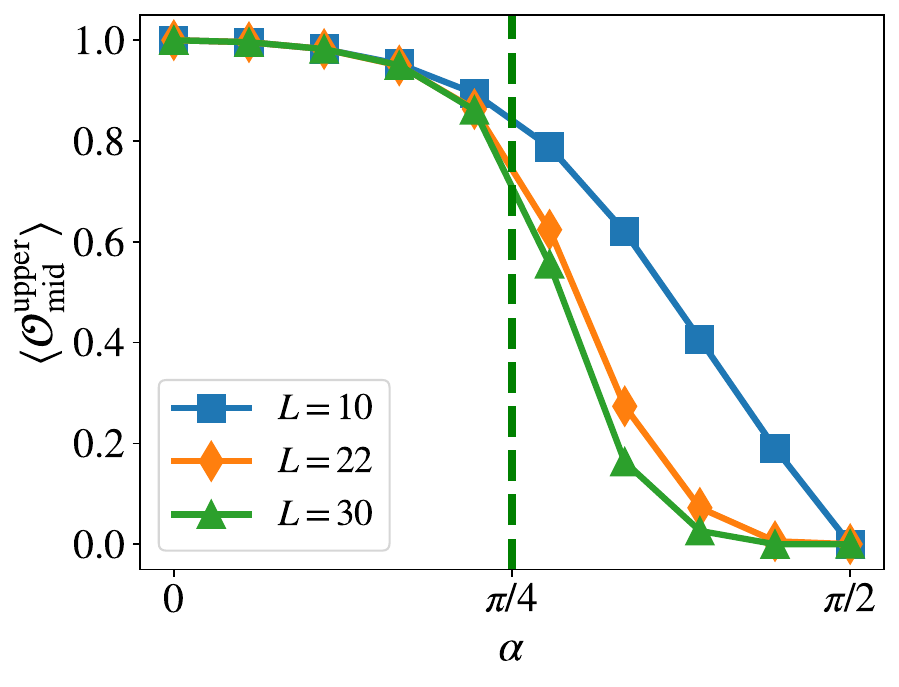}
\end{center}
\caption{\label{fig:sop_sym}The SOP, $\langle \mathcal{O}^{\mathrm{upper}}_{\mathrm{mid}} \rangle$, of the upper path plotted for the hourglass resource states used for the MBQC teleportation protocol (without an input qubit) in the presence of symmetric noise. The phase transition at $\alpha_c = \pi/4$ is sharp in the limit $L\to \infty$, indicating that the computational power for rotations disappears for $\alpha> \alpha_c$.}
\end{figure}

The results for the upper path are shown in Fig.~\ref{fig:fig_fidelity_up_sym}. For each $\alpha$ we generated a resource state. Each point corresponds to 25 random input states. The average over the random states and the runs converges to the fidelity of the MBQC protocol, which is the success probability to teleport random input state. Our MPS calculations show that the fidelity using unperturbed paths is unity for all converged MPS data, except the unentangled point $\alpha=\pi/2$. The interesting fact that perfect teleportation persists in the presence of symmetric perturbations has been shown in Ref.~\cite{azses2020identification}.  However, we note that computational power, which  includes single qubit rotations, does degrade with $\alpha$. This degradation is encoded in the SOP which is an SPT order parameter, as we now discuss.

\begin{figure}[t]
\begin{center}
\includegraphics[width=.8\linewidth]{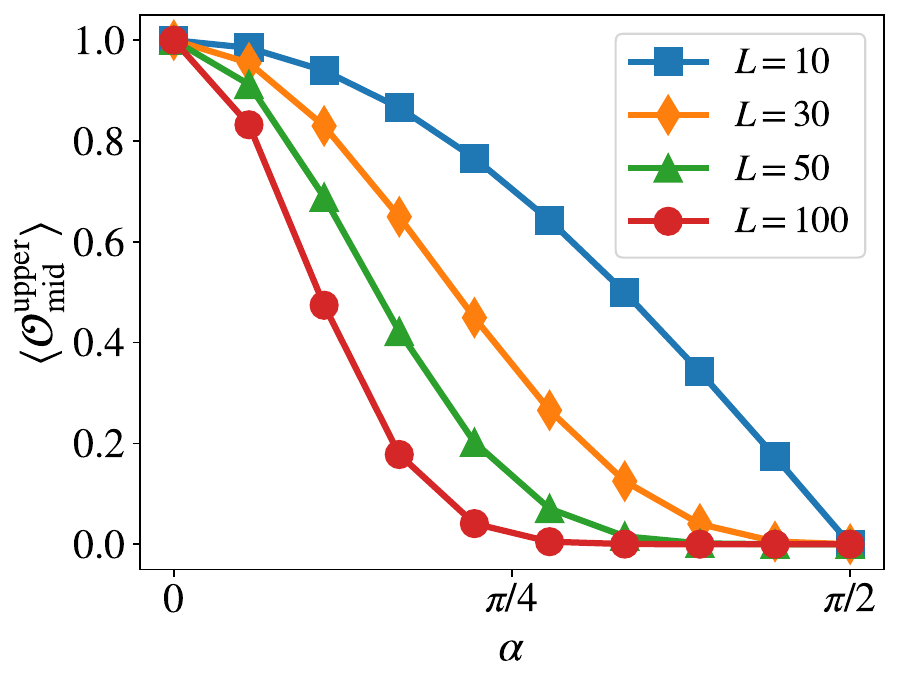}
\end{center}
\caption{\label{fig:sop_nonsym}The SOP, $\langle \mathcal{O}^{\mathrm{upper}}_{\mathrm{mid}} \rangle$, of the upper path plotted for the hourglass resource states used for the MBQC teleportation protocol (without an input qubit) in the presence of non-symmetric noise. As $L$ increases, the computational power diminishes faster and completely vanishes for smaller $\alpha$. This indicates that in the limit $L \to \infty$ we may have no computational power even for any $\alpha>0$ as it seems to vanish exponentially fast.}
\end{figure}

Let us consider the SOP for a particular path. Specifically, we consider
\be
\label{eq:sop}
\mathcal{O}^{\mathrm{upper}}_{\mathrm{mid}} = Z_{(l-1,0)}Z_{(l-1,1)} \left( \prod_{L/4>r\geq \lfloor L/8 \rfloor} Y_{L/2-2r,0} \right) X_O,
\ee
where $X_O$ acts on the output (rightmost) qubit and $l=L/2-2 \lfloor L/8 \rfloor$.  Let us discuss the physical meaning of this SOP. It consists of a product of stabilizers $K_j$ from the middle of the chain at site $l$, and along the upper path, all the way to and including the output qubit. A finite value of  $\mathcal{O}^{\mathrm{upper}}_{\mathrm{mid}} $ is an SPT order parameter, which can be path dependent. 

The results are given in Fig.~\ref{fig:sop_sym}. As we can see for the unperturbed state at $\alpha=0$ we have a unit SOP, $\mathcal{O}^{\mathrm{upper}}_{\mathrm{mid}}=1$, which decreases with $\alpha$. A similar earlier study for a single chain~\cite{RAUSSENDORF2023} observes a phase transition at intermediate $\alpha$. We expect that our system shows a similar phase transition at an intermediate $\alpha$ although the system's length that we approached $L \le 30$ do not allow to determine its location. For the Hamiltonian Eq.~(\ref{eq:Hy}) the perturbation is symmetric, hence the SOP does not depend on the path.

\subsection{Quasi-global non-symmetric perturbation}
Next we consider the resource state derived from the Hamiltonian 
\be
\label{eq:Hz}
H_{\text{Z}}= -\cos{\alpha} \sum_{j=1}^{L} K_j  -\sin{\alpha}  \sum_{i=1}^n (Z_{i,0}+Z_{i,1}).
\ee 
For the ground state of this Hamiltonian, the SOP Eq.~\ref{eq:sop} is plotted  in Fig.~\ref{fig:sop_nonsym}.  As the system size grows the SOP diminishes to $0$ very fast, indicating that the perturbation destroys the inherent computational power of the resource state even for small $\alpha$. Also in this case, the SOP does not depend on the selected path, and is diminished with $\alpha$ for both upper lower paths paths.

\subsection{Non-symmetric perturbation on one path} 
\begin{figure}[t]
\centering
\includegraphics[width=0.8\linewidth]{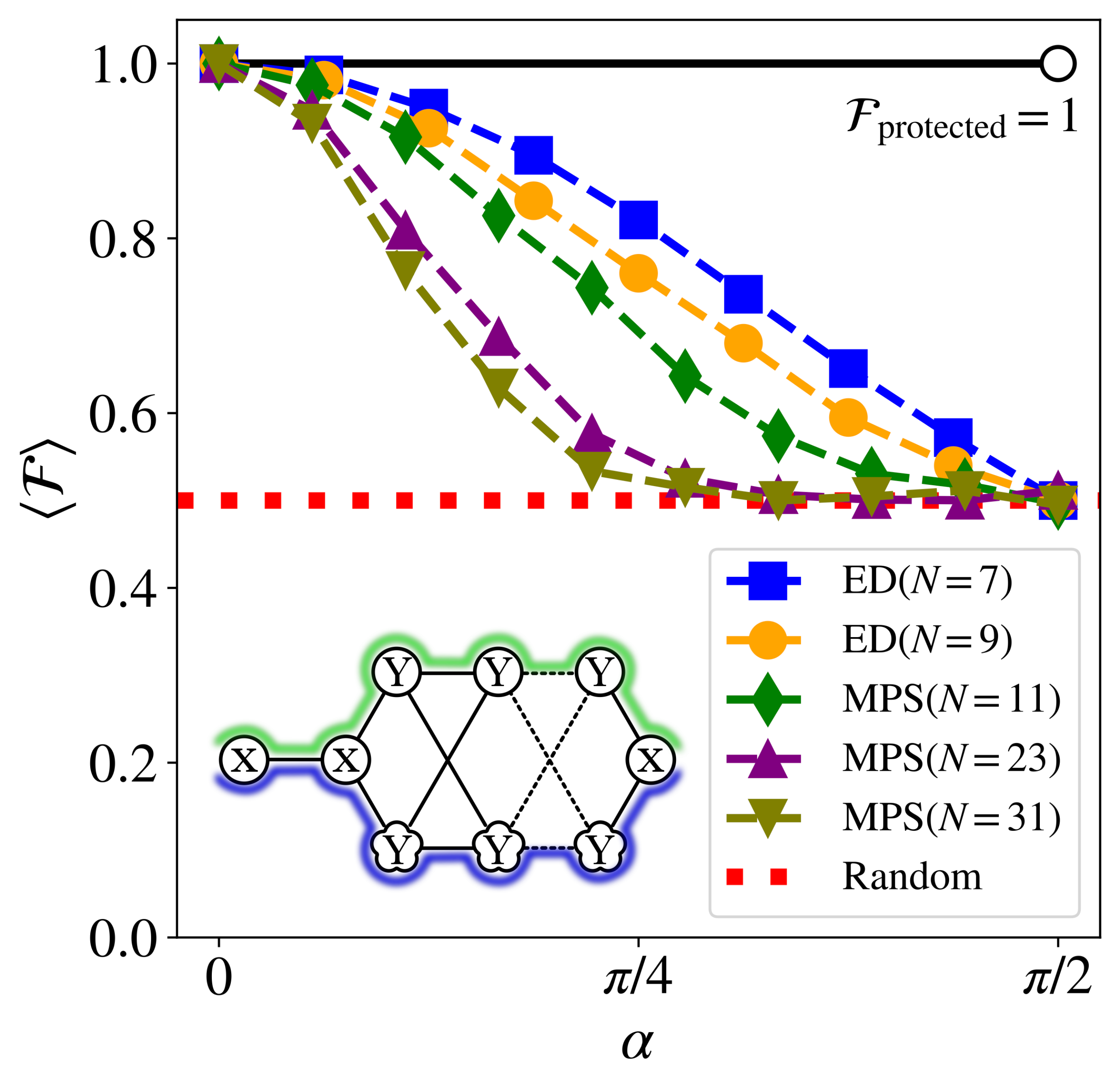}
\caption{\label{fig_hourglass_sim_graph} 
Teleportation fidelity versus $Z$-perturbation strength in the hourglass graph [Eq.~\eqref{eq:H_hourglass_non_symmetric}] averaged over several different input qubit orientations. The solid line shows perfect teleportation along the upper path obtained for all system lengths for all but the $\alpha=\pi/2$ point. The symbols, obtained using exact diagonalization (ED) and MPS, are for various graph lengths, 
$N=2n+3$ (See Fig.~\ref{fig_hourglass_graph}), along the lower path. Teleportation along the perturbed path degrades with increasing graph length.  For the ED we used 100 random input states in the Bloch sphere. For the MPS calculation we used 75 random states in the XY plane as input, each sampled by 100 measurements, see Appendix~\ref{app:mps_symmetric} 
for description of the MPS simulations.
}
\end{figure}

Finally, we consider the Hamiltonian
\bea
\label{eq:H_hourglass_non_symmetric}
H_{\text{Z}}^{\text{lower}}= -\cos{\alpha}\sum_{j=1}^L K_j-\sin{\alpha}\sum_{i=1}^n Z_{i,1}, 
\eea
where the non-symmetric perturbations act only on the lower sites of the hourglass graph.  The black line in Figure~\ref{fig_hourglass_sim_graph} plots perfect fidelity, which we explicitly obtained for $\alpha<\pi/2$, arising from teleportation along the upper path, which supports the exact symmetries Eq.~(\ref{hourglasssym}) with $k=0$.  The remaining lines show degradation in teleportation fidelity when using the lower path arising from broken string order. These classical simulations show that large $n$ still permits perfect transmission along the graph in spite of large numbers of non-symmetry preserving errors in the graph. 

\section{SPT order and degeneracies in the reduced Density Matrices}
\label{app:reduced_dm}
In this section, we show that the error protection results from  protected degeneracies in the entanglement spectrum, which are characterized by quantum numbers of the symmetry which survives in the presence of errors.

We consider pure graph states described by density matrix $ \rho^G=|\Psi^G\rangle\langle\Psi^G|$. To identify SPTO in these graph states, we perform a
Schmidt decomposition into regions $A$ and $B$,
\begin{align}
|\Psi^G\rangle=\sum_{\nu} \sqrt{\lambda_{\nu}}|\psi_{\nu}^A\rangle|\psi_{\nu}^B\rangle.
\end{align}
This allows to construct and diagonalize the reduced density matrices $\rho^G_{\text{A}}$, having eigenvalues $\lambda_{\nu}$, 
\begin{align}
    \rho_A^G={\rm{Tr}}_B \rho^G=\sum_{\nu}\lambda_{\nu}|\psi_{\nu}^A\rangle\langle\psi_{\nu}^A|.
\end{align}
As we now discuss for the chain and diamond graphs, the  symmetry operators acting on the subsystem for each graph can also be diagonalized.  The resulting reduced density matrix is then block diagonal~\cite{azses2020identification,PhysRevB.104.L220301} in terms of these symmetries.

\subsubsection{
Chain Graph}
Let us explicitly diagonalize the reduced density matrix of the 6 vertex chain graph defined by $\vert \Psi_G (0)\rangle$.  We first trace out vertices 1, 2, and 3  of the chain (region $B$) to obtain the reduced density matrix for region $A$, and obtain
\begin{align}
 \rho_{A}^C=\sum_{\nu=1}^2 \lambda_{\nu}^C|\psi_{\nu}^C\rangle\langle\psi_{\nu}^C|,
 \label{eq_chain_rdm}
\end{align}
where
\begin{align}
|\psi_1^C\rangle&=\frac{1}{\sqrt{2}}(|+\rangle^z_4|-\rangle^y_5|+\rangle^z_6+|+\rangle^z_4|+\rangle^y_5|-\rangle^z_6 \nonumber), \\
|\psi_2^C\rangle&=\frac{1}{\sqrt{2}}(|-\rangle^z_4|+\rangle^y_5|+\rangle^z_6+|-\rangle^z_4|-\rangle^y_5|-\rangle^z_6
\label{eq_chain_rdm_states}),
\end{align}
where $\vert \pm \rangle^x_i$, $\vert \pm \rangle^y_i$, and $\vert \pm \rangle^z_i$ denote the eigenstates of the Pauli matrices $X_i$, $Y_i$, and $Z_i$, respectively.  
Equations~\ref{eq_chain_rdm} and \ref{eq_chain_rdm_states} show, by explicit diagonalization, that $\rho_{A}^C$ has degenerate eigenvalues, $\lambda_1^C=\lambda_2^C=1/2$.  

We now show that the eigenstates of $\rho_{A}^C$ are also eigenstates of the reduced chain symmetries.  As discussed in Sec.~\ref{se:2}, the chain graph has symmetry generators $s_{135} = X_1 Y_3 Y_5 Z_6$ and $s_{246} = Z_1 Y_2 Y_4 X_6$.  By tracing out the vertices in region $B$ we obtain reduced symmetries  $Y_5 Z_6$ and $Y_4 X_6$.  We note that the reduced symmetries do not commute.  Furthermore: 
\begin{align}
Y_5 Z_6 |\psi^C_{r}\rangle&=(-1)^{r}|\psi^C_{r}\rangle, \nonumber \\
Y_4 X_6 |\psi^C_{r}\rangle &= (-1)^{r+1}i|\psi_{3-r}^C\rangle, 
\end{align}
for $r=1$ and $2$.  

We  see that we can simultaneously diagonalize  
$\rho_{A}^C$ and either $\text{Tr}_{\text{B}}[s_{246}]$ or $\text{Tr}_{\text{B}}[s_{135}]$. The non-commutativity of the latter two, along with their commutativity with $\rho_A^C$, directly implies the degeneracies of eigenvalues. This degeneracy is a property of the SPT phase  protected by the $(\mathbb{Z}_2)^2$ symmetry.

\subsubsection{Diamond Graph }

Now consider the 
diamond graph.  Tracing out vertices 1,2,3 in region $B$, we obtain the reduced density matrix for region $A$, $\rho_{A}^D$.  We diagonalize the reduced density matrix and find
\begin{align}
      \rho_{A}^D=\sum_{\nu=1}^4 \lambda_{\nu}^D|\psi_{\nu}^D\rangle\langle\psi_{\nu}^D|,
  \end{align}
where 
  \begin{align}
 |\psi_1^D\rangle&=|+\rangle^z_4|+\rangle^z_5|+\rangle^x_6, \nonumber \\
    |\psi_2^D\rangle&=|+\rangle^z_4|-\rangle^z_5|-\rangle^x_6, \nonumber \\
    |\psi_3^D\rangle&=|-\rangle^z_4|+\rangle^z_5|+\rangle^x_6, \nonumber \\
      |\psi_4^D\rangle&=|-\rangle^z_4|-\rangle^z_5|-\rangle^x_6.
  \end{align}
We  see four degenerate eigenvalues in the reduced density matrix, $\lambda_1^D=\lambda_2^D=\lambda_3^D=\lambda_4^D=1/4$.
  
We now show that the eigenstates of $\rho_{A}^D$ are also eigenstates of the reduced chain symmetries.   Recall from  Sec.~\ref{se:3} that the diamond graph has the symmetry generators  
\bea
s_{135}&=&X_1 X_3 X_5 Z_6, \nonumber \\
s_{246}&=&Z_1 X_2 X_4 X_6, \nonumber \\
s_{145}&=&X_1 X_4 X_5 Z_6, \nonumber \\ 
s_{236}&=&Z_1 X_2 X_3 X_6.\nonumber
\eea
Tracing out vertices 1,2, and 3 leaves the reduced symmetry generators $ X_5 Z_6$, $X_4 X_6$, $X_4 X_5 Z_6$, and $X_6$.  We now have two symmetry groups since $[X_4 X_5 Z_6, X_6]\neq0$ and $[X_5 Z_6, X_4 X_6]\neq0$.  We can construct the eigenstates of the reduced symmetries
\begin{align}
X_6|\psi_r^D\rangle  &=(-1)^{r+1}|\psi_r^D\rangle,
\nonumber \\
X_4X_6 |\psi_r^D\rangle &=(-1)^{r+1}|\psi_{r+2~ {\rm{mod}} ~4}^D\rangle,
\nonumber \\
X_4 X_5 Z_6 |\psi_r^D\rangle &=|\psi_{5-r}^D\rangle,
\nonumber \\
X_5Z_6 |\psi_{2r-1}^D\rangle &=|\psi_{2r}^D\rangle.
\end{align}

We can simultaneously diagonalize  
$\rho_{A}^D$ and two commuting reduced string symmetries, for example $X_6$ and $X_4 X_6$, to obtain the 4 degenerate eigenstates. This corresponds to the $(\mathbb{Z}_2)^4$ symmetry.  Upon adding an error which lowers the symmetry, for example $\epsilon_{35}$, we are still left with two symmetries $s_{135}$ and $s_{246}$, which as discussed above, still preserve a degeneracy in the entanglement spectrum corresponding to the lower symmetry $(\mathbb{Z}_2)^2$ which is still sufficient for teleportation.

\section{Summary and Outlook}
\label{se:5}We constructed an error correction and detection protocol using non-local symmetries in graph states.  By showing that certain graph states contain enlarged $(\intz_2)^g$ SPT order we used the ensuing redundancy in symmetry to locate and correct errors.  The protocol was demonstrated to protect non-local measurement-based teleportation along a quantum channel in quantum devices for example graphs.  

Our work has implications for correcting errors in MBQC resource states on noisy near-term quantum computers where $ZZ$-crosstalk is non-trivial to correct using conventional error correction with local stabilizers.  Furthermore, our protocol can be extended beyond teleportation to correct other logical operations.  We also note that our work has applications to low-overhead correction of error-prone quantum networks.  

As a related concept, we note that a  recent work~\cite{eckstein2024robust} showed error protection of teleportation of a qubit encoded in a surface code.

\begin{acknowledgments}
We acknowledge support from ARO W911NF2010013.  ES acknowledges support from the European Research Council (ERC) Synergy funding for Project No. 951541, and the Israel Science Foundation, grant number 154/19.  RR is funded by NSERC.  VWS and ZQ acknowledge support from AFOSR (FA2386-21-1-4081, FA9550-23-1-0034, FA9550-19-1-0272) and ARO W911NF2210247. We acknowledge use of the IBM quantum cloud experience and IonQ for this work. The views expressed are those of the authors and do not reflect the official policy or position of IBM, the IBMQ team, or IonQ.
\end{acknowledgments}

\appendix

\begin{figure}
\begin{center}
\includegraphics[width=0.3\textwidth]{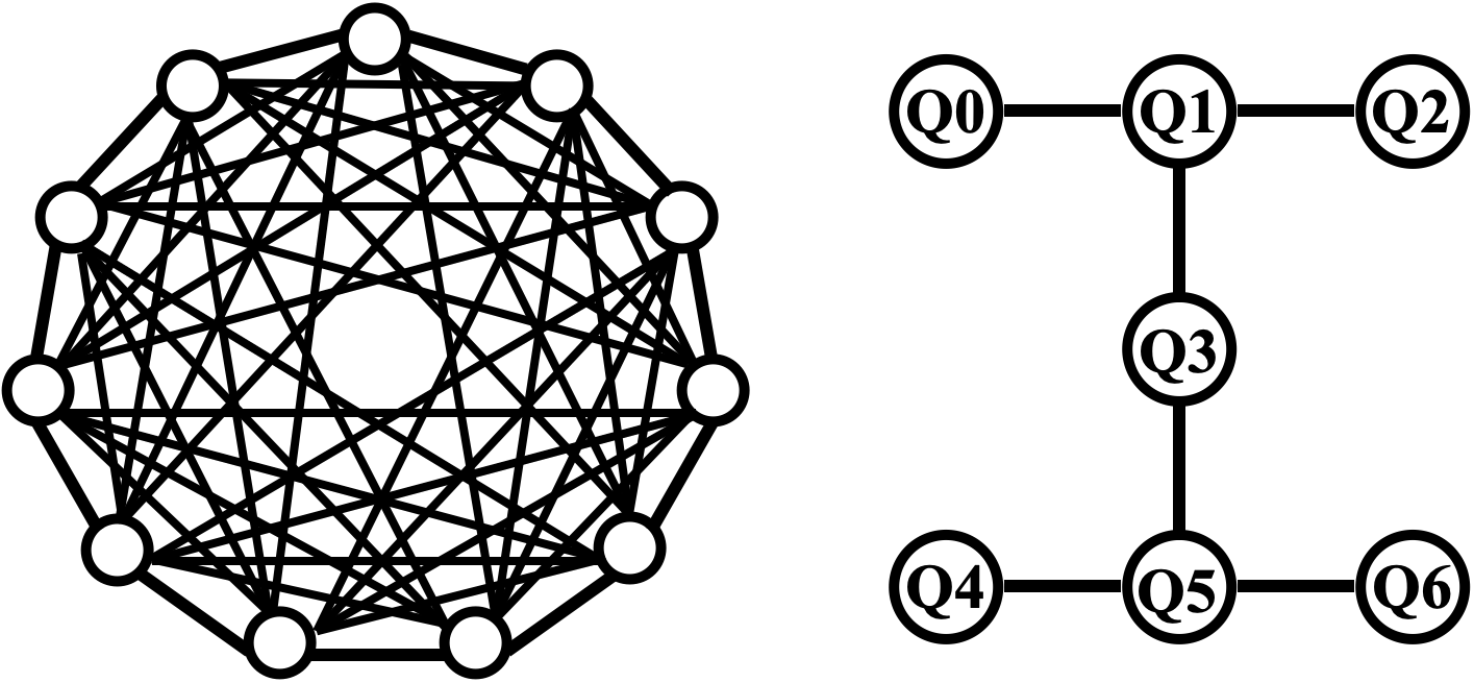}
\end{center}
\caption{ \emph{Quantum Device Geometries}:
Schematics for IonQ Harmony (left) and IBMQ Nairobi (right) device qubit connectivities. The circles represent qubits and the lines indicate two-qubit gates. Gate fidelities for IBMQ and IonQ are reported in Tables~\ref{table_supp_ibm_real_device} and ~\ref{table_ionq_real_device}, respectively.}
\label{fig_device_geometry}
\end{figure}

\section{\label{app:circuit_device}Quantum Circuits and Quantum Device Parameterization}

The diamond and hourglass graph states discussed in the main text were implemented on two different quantum devices, IBMQ Nairobi and IonQ Harmony.  Demonstrations of teleportation of input states across the graphs were then performed.  Fig.~\ref{fig_device_geometry} shows schematics of the qubit connectivity for both devices.

\begin{figure}
\begin{center}
\includegraphics[width=0.4\textwidth]{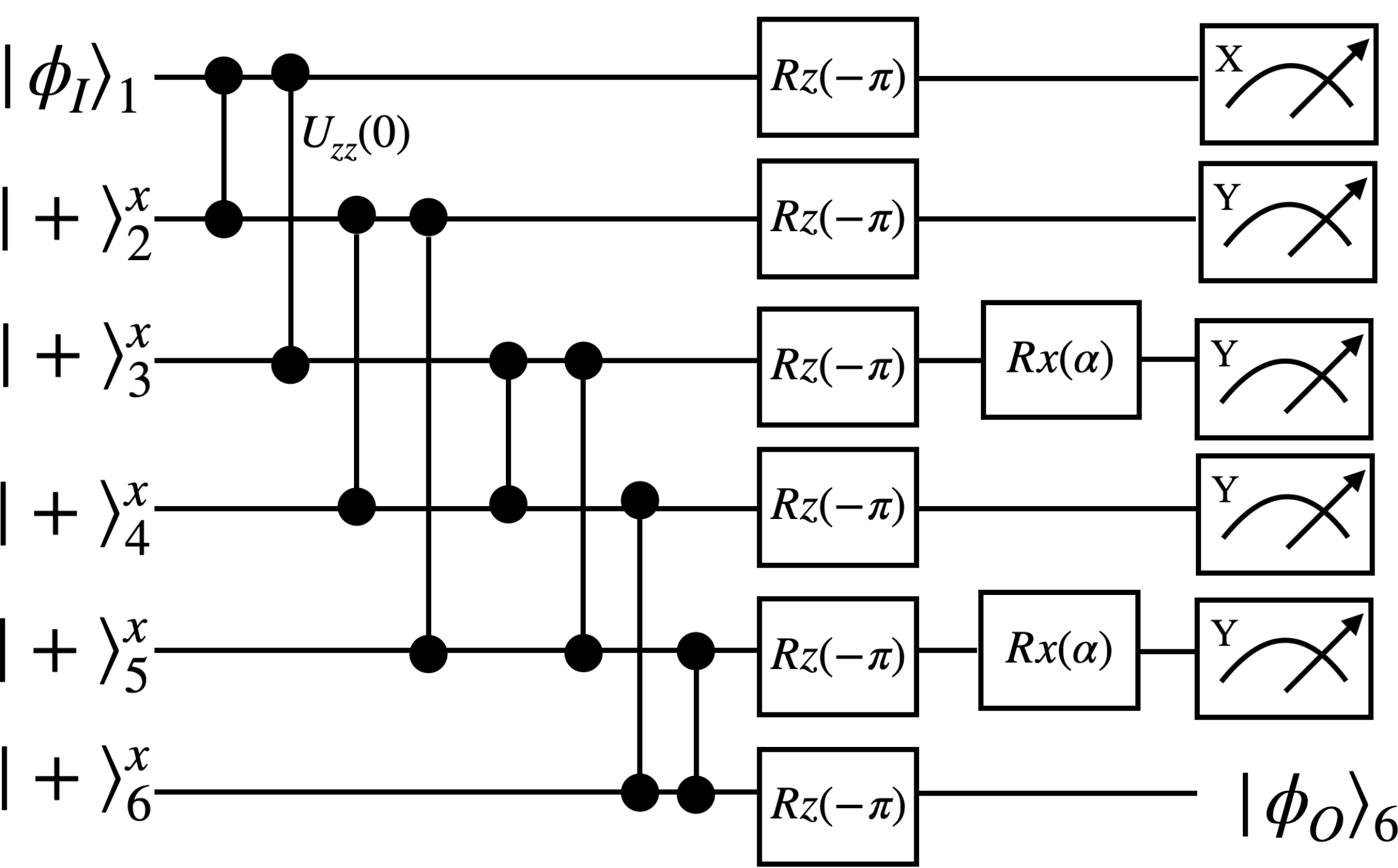}
\end{center}
\caption{\emph{Hourglass Graph Circuit Diagram}: Same as in Fig.~\ref{fig_circuit_diamond} but used for the hourglass graph with final qubit measurements along the $x$-direction. Single qubit $x$-rotations are then applied, where $R_{x}(\alpha) =e^{-i\alpha X/2}$.}
\label{fig_circuit_hourglass}
\end{figure}

Figures~\ref{fig_circuit_diamond} and \ref{fig_circuit_hourglass} show the quantum circuits used to prepare the graph states.  Initialization along the qubit-$x$ direction is followed by entangling Ising gates. The $z$ and $x$-rotations were performed to set up the appropriate measurement bases.  The final measurements implement teleportation along the graph states, from qubit 1 to qubit 6. 

The real quantum device data presented in the main text are averages obtained using 10000 (100) shots on IonQ for the diamond (hourglass) graph, and 8192 shots on IBMQ.  The standard deviation in Gaussian fits are about 0.05 and 0.1 for IonQ and on IBMQ, respectively.

\begin{table*}
\centering
\scriptsize
\begin{tabular}{l|lllllllllllll}
\hline\hline
Qubit & T1 ($\mu$s) & T2 ($\mu$s) & F (GHz) & A (GHz) & RA ($\times 10^{-2}$)  & M0P1 & M1P0 & RL (ns) & ID ($\times 10^{-4}$)  & SX ($\times 10^{-4}$) & PX ($\times 10^{-4}$)  & CN ($\times 10^{-3}$)  & GT (ns) \\
\hline
Q0 &  107.5 & 28.11 & 5.26 & -0.33983 & 2.140 & 0.0284 & 0.0144 & 5560.889 & 3.282 & 3.282 & 3.282 & 0-1:7.243 & 248.889 \\
Q1 & 137.77 & 67.82 & 5.17 & -0.34058 & 2.310 & 0.0294 & 0.0168 & 5560.889 & 2.873 & 2.873 & 2.873 & 1-3:1.078 & 270.222 \\
Q2 & 85.88  & 96.75 & 5.274 & -0.3389 & 4.010 & 0.0626 & 0.0176 & 5560.889 & 2.406 & 2.406 & 2.406 & 2-1:7.034 & 391.111 \\
Q3 & 77.05 & 48.93 & 5.027 & -0.34253 & 6.350 & 0.088 & 0.039 & 5560.889 & 5.493 & 5.493 & 5.493 &  &  \\
Q4 & 111.79 & 75.6 & 5.177 & -0.34059 & 2.050 & 0.031 & 0.01 & 5560.889 & 2.625 & 2.625 & 2.625  & 5-4:6.453 & 277.333 \\
Q5 & 119.8 & 18.5 & 5.293 & -0.34053 & 2.280 & 0.0294 & 0.0162 & 5560.889 & 3.202 & 3.202 & 3.202 & 5-3:2.101 & 241.778 \\
Q6 & 113.77 & 114.03 & 5.129 & -0.34044 & 2.080 & 0.035 & 0.0066 & 5560.889 & 2.325 & 2.325 & 2.325  & 6-5:7.750 & 305.778 \\
\hline\hline
\end{tabular}
\caption{Calibration data of the IBM device backend, {\it ibm\_nairobi}, used to generate the real-device simulation date. Abbreviations are defined as follows: F = frequency, 
A = anharmonicity, 
RA = readout assignment error, 
M0P1 = probability of measurement 0 and preparation 1, 
M1P0 = probability of measurement 1 and preparation 0, 
RL = readout length, 
SX = $\sqrt{X}$ error, 
PX = Pauli-X error, 
CN = CNOT error, 
GT = gate time.}
\label{table_supp_ibm_real_device}
\end{table*}

\begin{table*}[ht]
\centering
\scriptsize
\begin{tabular}{l|lllllllllllll}
\hline\hline
Qubit & T1 (s) & T2 (s) & GT-1Q(s) & GT-2Q(s) & RT(s) & RS(s)& F-1Q & F-2Q & SPAM\\
\hline
Q0-10(Mean) & 10000 &0.2 & 0.00001 & 0.0002 &  0.00013 &0.0002 & 0.9986 &0.9726 & 0.99752\\
\hline\hline
\end{tabular}
\caption{Calibration data of the IonQ device backend, used to generate the real-device simulation data. Abbreviations are defined as follows: GT-1Q= single qubit gate time, GT-2Q= two qubit gate time, RT=readout time, RS=reset time, F-1Q= single-qubit gate fidelity, F-2Q=two-qubit gate fidelity, SPAM=state preparation and measurement fidelity}
\label{table_ionq_real_device}
\end{table*}

The data are significantly impacted by noise  beyond the unitary noise sources discussed in the main text.   These noise sources suppress the fidelity beyond our analysis.   Tables~\ref{table_supp_ibm_real_device} and  \ref{table_ionq_real_device} show device parameters from the quantum devices used to create the data shown in the main text.

We model the noise on the IonQ device.  
Table~\ref{table_ionq_real_device} shows that two-qubit gate fidelity is lowest.  The native two-qubit gates in the IonQ device \cite{Wright2019} are Mølmer-Sørenson Ising-based gates.  We model non-unitary perturbations using two-qubit gate depolarizing noise defined in the Qiskit noise model \cite{Qiskit}.  We use the two-qubit depolarizing channel as a proxy for cumulative error due to a variety of non-unitary noise sources \cite{Wu2018,Wright2019}.  

\begin{figure}
\begin{center}
\includegraphics[width=0.5\textwidth]{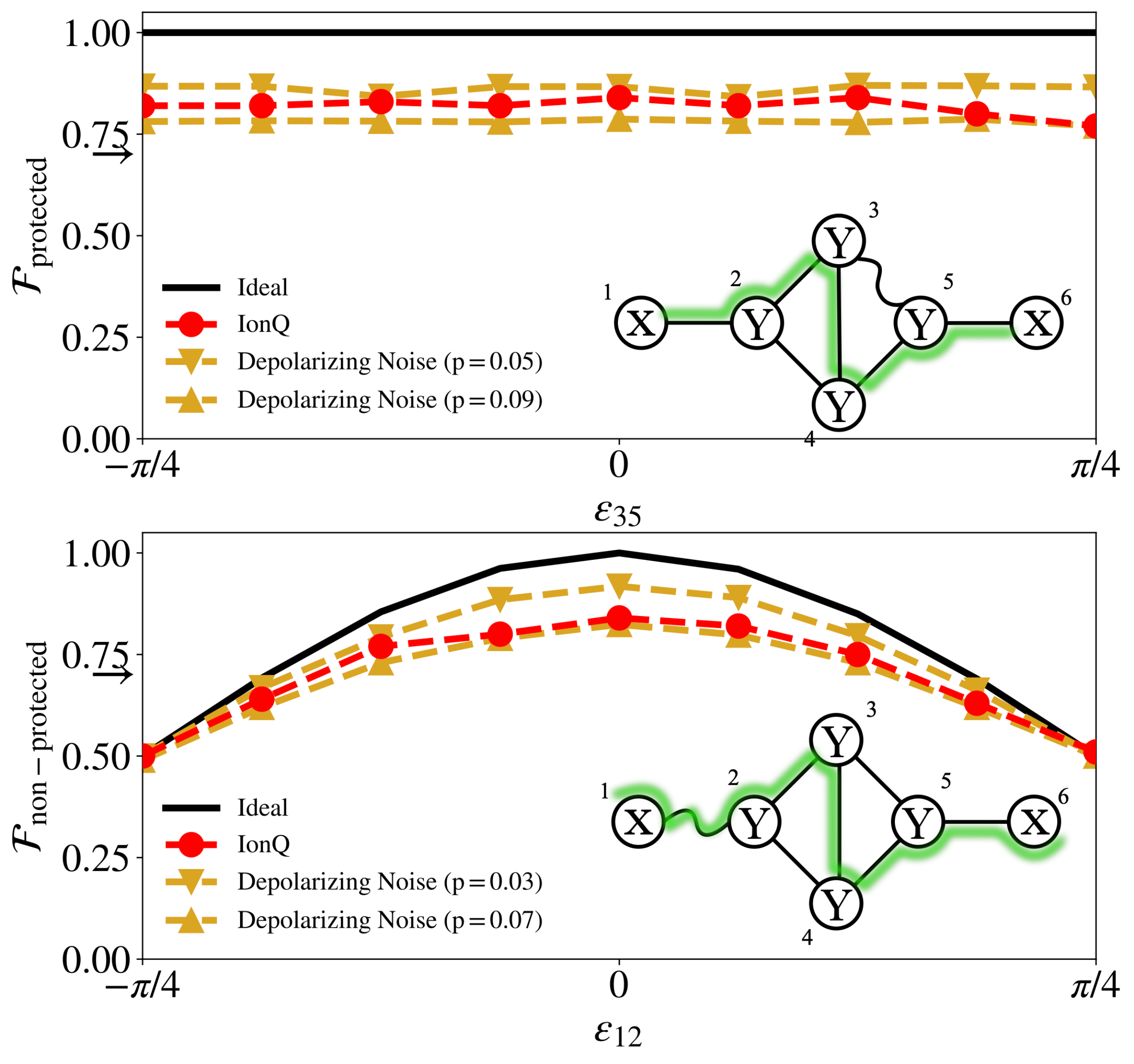}
\end{center}
\caption{
The same as the IonQ data in Fig.~\figfidelitymachinesz in the main text but with classical noise-model simulations included (triangular symbols).  The classical noise-model simulations include a calculation of the fidelity with depolarizing noise applied to two-qubit gates used to build the graph state.  The fidelity is measured using the initial qubit aligned along the $+z$ direction.  $p$ is the probability of the two-qubit depolarizing noise \cite{Qiskit}.  The qualitative accuracy between the noise model and the real machine data indicate that two-qubit depolarizing noise offers a good model for non-unitary perturbations found on IonQ. 
}
\label{fig_depolarizing_expeirment}
\end{figure}

Figure~\ref{fig_depolarizing_expeirment} compares real device teleportation demosntrations with simulation.   In the classical simulations, we choose two different two-qubit depolarizing noise probabilities to bound the quantum device data.  We see that the real device data points (red circles) lie between simulator results (upward and downward triangles) which captures the qualitative effect of noise seen on the real device.  We therefore conclude that two-qubit depolarizing noise serves as a good model for the dominant noise channel which, in turn, lowers the teleportation fidelity for IonQ Harmony from the ideal values discussed in the main text.

\section{\label{app:mbqc_path_calibration}MBQC Protocol Path Calibration}
In this appendix we describe the calibration process of finding the correct path to use and how to find on what sites the errors occur. We focus on 1-qubit error here as 2 errors may block both paths, reducing the fidelity of the teleportation. We further assume, as in other noisy quantum device error mitigation techniques, that the errors do not change drastically from the calibration process to the actual hardware.

Let us focus on the case of Fig.~\fighourglassgraph in the main text. Here we have the upper path and the lower path, where one path is broken but we do not know which. To find the broken path we teleport a simple known state such as $\ket{0}$ or $\ket{+}$, as those do not require tomography or different quantum measurements for the paths, and use the byproduct operators of both paths on the output to get two bits $s_{\mathrm{out}}^{\mathrm{upper}}$, $s_{\mathrm{out}}^{\mathrm{lower}}$, each for its respective path. In the case of no errors, both should be $0$, indicating perfect fidelity, however, as one path is erroneous the fidelity will not be perfect. On a real quantum computer, it may be that both paths contain some errors, but one path has decreased fidelity with respect to the other one ({\it i.e.} $90\%$ vs $60\%$ fidelity). Hence, we calibrate the quantum computer to the best path using the average fidelity over the simple quantum states. Additionally, we may use tomography and sample input states randomly to have better sampling. Therefore, in the case of a single erroneous path the calibration method always finds the correct path.

In the case that the input state is unknown and only one sample is given, one may use a graph similar to the one in Fig.~\fighourglassgraph, but with 3 rows instead of 2. In this case, if only one error is present, or only one path is broken, one can use the byproducts of each path to obtain three distinct results such that two results will be correct and one will be wrong as we assume only one path is perturbed. Using the majority rule, one obtains the correct result. Furthermore, the perturbed path is detected with only a single measurement. One may find it analogous to the classical repetition code with 3 bits, where one is able to detect and correct one error using majority rule.

\section{\label{app:mps_symmetric}Matrix Product State Simulations}
In this appendix we describe our explicit MPS procedure to perform teleportation through the MPS ground states of Hamiltonians Eq.~(\ref{eq:Hy}), (\ref{eq:Hz}), and (\ref{eq:H_hourglass_non_symmetric}).

First, the ground state for these Hamiltonians acting on $L=2+2n$ sites, excluding the input qubit, is constructed for any $\alpha$ using MPS which is found to be an excellent approximation to the true ground state. The resulting ground state is then used as the resource state for our MBQC protocol as follows:
\begin{enumerate}
    \item Preparation: We couple the obtained MPS states of $L=2n+2$ qubits to the additional input state. The input state is chosen randomly in the XY plane on the Bloch sphere, with a random azimuthal angle $\phi$. This input state is then entangled with a CZ gate with the left most qubit $m$, see Fig.~\ref{fig_hourglass_graph}.
    \item Measurement: We first apply a rotation on every qubit in the MPS according to the desired measurements, i.e.,  for $X$ measurements we apply an $H$ gate, and for $Y$ measurements we apply $HS^\dagger$. The output is not being acted on. We then sample the MPS, excluding the output, in the computational basis by applying the projectors according to the measurements probabilities. This results in measurement results $s_i=0,1$ for the qubit $i$, where $s_{\mathrm{in}}$ is the result for the input qubit.
    \item The output is measured in two steps. First we rotate around $Z$ by an angle $(-1)^{1+l} \phi$, which corrects the  $X$ byproduct operator, where  the sign is path dependent $l_{\mathrm{upper}}=s_m +\sum_k  s_{(2k,0)}, \ l_{\mathrm{lower}}=s_m + \sum_k  s_{(2k,1)}$. Then, a simple measurement in the $X$ direction follows by applying $H$ and the result is $s_{\mathrm{o}}$.
    \item We expect to get $s_{\mathrm{o}}=0$, but that is only true if there were no $Z$ byproduct operators that flip the result (in the $X$ basis). Thus, a flip is needed in the case of a $Z$ byproduct operator. This depends on the path, hence, we flip if $l=1 \pmod 2$, where $l_{\mathrm{upper}}=s_{\mathrm{in}} +\sum_k  s_{(2k-1,0)}, \ l_{\mathrm{lower}}=s_{\mathrm{in}} +\sum_k  s_{(2k-1,1)}$ are the $l$ for their respective paths. After flipping, the result is $s_{\mathrm{o}}=0$ if the MBQC teleportation protocol was successful. In Fig.\ref{fig:fig_fidelity_up_sym} (Fig.~\ref{fig_hourglass_sim_graph}) we performed 10 (100) measurements in order to compute the fidelity for each of the 25 (75) random input states.
\end{enumerate}

\section{\label{app:mps_non_symmetric1}Non-Polar Input State and Quantum Tomography}

\begin{figure}
\begin{center}
\includegraphics[width=\linewidth]{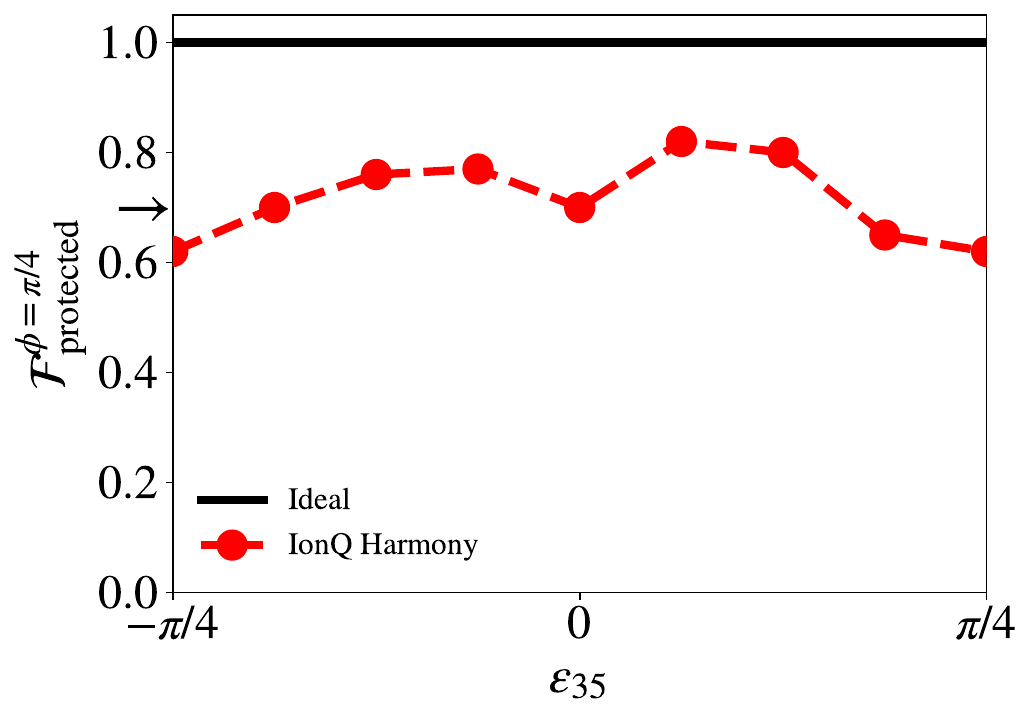}
\end{center}
\caption{\label{fig:pi_4_ionq} The same as the top panel of Fig.~\ref{fig_fidelity_machines_z} but for an input state that that is at an angle $\pi/4$ in the $XY$ plane.  The black line shows the ideal case and the red circles are data taken on the IonQ device. }
\end{figure}

In this appendix, we describe the methodology for performing quantum tomography on a randomly selected input state within the Bloch sphere.  We illustrate this process using a Diamond graph example where the input state resides in the $XY$ plane.

To teleport a qubit prepared in the $XY$ plane, we implement a modified protocol.  First, we include a Hadamard gate to the output qubit.  We then must update the byproduct operator.  We must also perform repeated measurements on the output qubit along all three directions.  Binning measurement outcomes in all three directions allows us to reconstruct the density matrix of the output qubit, $\rho_m$, in the usual manner.  We then construct the teleportation fidelity to be: $\text{Tr}(\rho_{in}\rho_{m})$, where $\rho_{in}$ is the density matrix of the input qubit.

Fig.~\ref{fig:pi_4_ionq} plots results using this protocol.  The graph state and perturbations are chosen to be same as the top panel in Fig.~\ref{fig_fidelity_machines_z} but with the input state at an angle $\pi/4$ in the $XY$ plane and the fidelity measured to be $\text{Tr}(\rho_{in}\rho_{m})$.  The data were taken on the IonQ device.  The fidelities are lower than in the main text because the added tomography steps add errors.

\bibliography{references} 

\end{document}